\documentclass[%
 amsmath,amssymb,
 aps,
pra,
]{revtex4-2}

\usepackage{slashed}
\usepackage{xcolor}
\usepackage{cancel}
\usepackage{graphicx}
\usepackage{subfig}
\usepackage{soul}
\usepackage{xcolor}
\usepackage{dcolumn}
\usepackage{bm}
\usepackage{enumerate}

\begin{document}


\title {Dimensional Regularization in Quantum Field Theory with Ultraviolet Cutoff}

\author{Durmu{\c s} Demir$^1$, Canan Karahan$^2$, Ozan Sarg{\i}n$^1$}
 \affiliation{$^1$ Sabanc{\i} University, Faculty of Engineering and Natural Sciences, 34956 Tuzla, {\.I}stanbul, Turkey\\
 $^2$ National Defence University, Turkish Naval Academy, Department of Basic Sciences, 34942 Tuzla, {\.I}stanbul, Turkey}

\date{\today}

\begin{abstract}
In view of various field-theoretic reasons,  
in the present work, we study the question of if the usual dimensional regularization can be extended to quantum field theories with an  ultraviolet cutoff (Poincare-breaking scale) in a way preserving all the properties of the dimensional regularization.  And we find that it can indeed be. The resulting extension gives a framework in which the power-law and logarithmic divergences get detached to involve different scales. This new regularization scheme, the detached regularization as we call it, enables one to treat the power-law and logarithmic divergences differently and independently.  We apply the detached regularization to the computation of the vacuum energy and to two well-known QFTs namely the scalar and spinor electrodynamics. As a case study, we consider  Fujikawa's subtractive renormalization in the framework of the detached regularization, and show its effectiveness up to two loops by specializing to scalar self energy. We discuss various application areas of the detached regularization. 
\end{abstract}

\maketitle

\section{Introduction}\label{sec:intro}

Quantum field theories (QFTs) develop divergences at the loop level. The ultraviolet (UV) divergences refer to infinities arising at large loop momenta. The methodology to deal with the divergences --  renormalization  -- requires the divergent and finite terms to be separated appropriately -- regularization. Various regularization methods have been proposed since the early days of the QFTs \cite{regularization}. The simplest and physically most intuitive regularization scheme is  the cutoff regularization in which QFTs are ascribed a {\it Poincare-breaking} UV edge. In this scheme, QFTs are endowed with a hard momentum cutoff  so that loop integrals run up to the cutoff not to the infinity \cite{cutoff1}. The effective QFT at the end can have quartic, quadratic and logarithmic dependencies on the UV cutoff \cite{ghp1}. The most conspicuous drawback of this scheme is that it breaks explicitly all the gauge symmetries in the QFT because each gauge boson acquires a mass proportional to the UV cutoff \cite{gauge-break1,gauge-break2,ccb1,ccb2}. (For clarity, one can consider the symmetric phase of the QFT in which none of the gauge symmetries is  broken spontaneously so that the entire gauge breaking comes from the UV cutoff.)

The cutoff regularization has a {\it Poincare-conserving} alternative: the Pauli-Villars regularization \cite{pauli-villars}. In this scheme, one  introduces a soft momentum cutoff $M$ (which is the mass of an auxiliary field) instead of the hard momentum cutoff $\Lambda$ (which is not a particle mass though it can be numerically equal to a particle mass). Practically, in Pauli-Villars regularization one makes the  replacement $(p^2-m^2)^{-1} \rightarrow (p^2-m^2)^{-1} - (p^2-M^2)^{-1}$ in the loop integration so that the propagator decreases faster as the loop momenta $p$ tends to infinity. The auxiliary particle provides a Poincare-conserving cutoff (with opposite statistics with respect to the actual particle of mass $m$). The Pauli-Villars regularization has an advantage over the cutoff regularization in that it leaves the theory gauge invariant at each order of perturbation theory thanks to the precise prescription of the use of the regulators \cite{ccb1}. The gauge invariance here is restricted to the Abelian case, which is what we are interested in the present work. The difficulty of finding a physical interpretation for the auxiliary fields is the main drawback of the Pauli-Villars regularization.

The cutoff-based regularizations above have two alternatives: Analytic regularization (changing the power of the loop propagator) and dimensional regularization (changing the dimension of the momentum space). The analytic regularization is  based on the concept of analytic continuation \cite{areg1,areg2,areg3}. The parameter which is exploited for this purpose is the one that is obtained by carrying the power of the denominator of the propagator to the complex plane. For instance, suppose we have the propagator for a particle of mass $m$ and four-momentum $p_{\alpha}$ which contains the factor $(p^2-m^2+i0)^{-1}$, what analytic regularization does is that it replaces the propagator with $(p^2-m^2+i0)^{-n}$, where the regulating parameter $n$ is a complex number in general. The crucial point is that $\mathrm{Re}(n)$ must be large enough so that the integrals are made to converge. After the result is obtained, which normally depends on the parameter $n$, it is analytically continued to the domain which contains the integral value of $n=1$. The divergences do all manifest themselves as simple poles in the limit $n \to 1$. This lets one to renormalize the QFT simply by subtracting the divergent pole terms. The finite parts which remain constitute the desired renormalized QFT. The analytic regularization respects the Poincare symmetry, however, as it is the case in the cut-off regularization, breakdown of the gauge symmetries is unavoidable.  
  
The dimensional regularization is the most popular regularization method due to its ability to preserve the local symmetries of the QFT such as the gauge symmetries \cite{dim-reg1,dim-reg2}. It has been successful all along in making precise predictions (higher loop calculations), which have been tested at the LEP,  LHC and other colliders.  The idea behind this method is to change the integration measure of the loop integrals by carrying the dimensionality $D$ of the measure to the complex domain. Let us illustrate this with a simple example and see how it helps. Consider a loop integral of the form
  \begin{equation}
      I=\int \frac{d^4 p}{(2\pi)^4} \frac{1}{(p^2+i0)[(p-q)^2-m^2+i0]} \,
  \end{equation}
  which is a four dimensional integral over the loop momentum $p$. In this form and in the limit of large loop momenta, namely $p \to \pm\infty$, this integral diverges. However, if we consider the same integral evaluated in three space-time dimensions 
  \begin{equation}
      I=\int \frac{d^3 p}{(2\pi)^3} \frac{1}{(p^2+i0)[(p-q)^2-m^2+i0]} \,
  \end{equation}  
we see that, in the limit $p \to \pm\infty$, it readily converges to a finite value. What this example teaches us is that unbounded loop integrals over four dimensional spacetime can be rendered convergent if it is possible to reduce the dimensionality of the momentum space \cite{dim-reg1,dim-reg2,dim-reg3,ccb1}. The dimensional regularization involves only logarithmic divergences since at the end of calculations one takes $D\rightarrow 4$. In essence, dimensional regularization isolates divergences via the poles of $1/(D-4)$ such that their subtraction via MS or $\overline{MS}$ schemes leads to renormalization of the underlying QFT. Independence of the bare parameters from $\mu$ gives rise to renormalization group equations and their solutions determine how various interactions  vary with $\mu$ (the scale of the experiments). This can be done up to any loop order. For instance, the fine structure constant varies with $\mu$ such that its value at $\mu=M_Z$ agrees with the LEP measurements. The same thing happens with the LHC measurements and, is expected to be the same also at the FCC and ILC if there is no new physics at their scales.

Having briefly discussed the four main regularization schemes in the literature, let us now go back to the cutoff regularization. Let us consider a QFT with a Poincare-breaking UV cutoff $\Lambda$ \cite{cutoff1}. The matter loops lead to three types of UV-sensitivites:
\begin{enumerate}
    \item Quartic sensitivities going like $\Lambda^4$,
    \item Quadratic sensitivities going like $\Lambda^2$,
    \item Logarithmic sensitivities involving $\log \Lambda$.
\end{enumerate}
These UV sensitivities can be exemplified by  a typical loop integral 
\begin{eqnarray}
I_n = \int  \frac{d^4 p}{(2\pi)^4} \frac{1}{(p^2-m^2+i0)^n}
\label{int-4d}
\end{eqnarray}
which is seen to diverge at large loop momenta for $n\leq 2$. It can be rendered finite (regularized) by cutting off all loop momenta above $\Lambda$ so that one gets the $\Lambda$-dependent finite results
\begin{eqnarray}
I_n(\Lambda^2) = \left\{\begin{array}{ll} \frac{i}{32 \pi^2} \Lambda^4 & \quad n=0\\ \\
\frac{-i}{16 \pi^2} \left(\Lambda^2- m^2 \log \frac{\Lambda^2+m^2}{m^2}\right) & \quad n=1\\ \\
\frac{i}{16 \pi^2} \left(-1 + \log \frac{\Lambda^2+m^2}{m^2}\right) & \quad n=2
\end{array}\right.
\label{int-4d-results}
\end{eqnarray}
corresponding, respectively, to the three types of the UV-sensitivities listed above. (As we deal with a renormalizable QFT, we do not consider non-renormalizable corrections of the form $1/\Lambda^n$  ($n>2$) or we assume that such corrections are absorbed in redefinitions of the QFT parameters.)  The power-law (quadratic and quartic) UV-sensitivities are local in that they live at the scale $\Lambda$. The logarithmic UV-sensitivities, on the other hand, extend in the entire range since the correction $M^2 \log \Lambda^2/M^2$ varies with the particle mass  $M$. Besides, the logarithmic UV-sensitivities set the beta functions and conformal anomalies in the QFT while the power-law divergences do not play such roles. In view of these differences, certain QFT studies have attempted to treat the power-law and logarithmic divergences differently and independently. One example is subtraction of the power-law divergences as in \cite{quadratic-logarithmic1,quadratic-logarithmic2,quadratic-logarithmic3}. Another example is induced gravity models in which $\Lambda^2$ leads to Newton's constant \cite{sakharov1,sakharov2}.  Yet another example is emergent gravity models in which $\Lambda^2$ is promoted to curvature (in reminiscence to promotion of vector boson masses to Higgs field) \cite{demir1,demir2,demir3}. In all such attempts, the hampering problem is  that both the logarithmic and power-law  divergences involve one and the same scale -- the UV cutoff $\Lambda$. They are thus not eligible for treating differently and independently.  In view of the examples above, however, we ask the crucial question:  Is it possible to find a new regularization scheme in which the power-law and logarithmic divergences are detached to involve independent scales in place of a single scale like $\Lambda$?
This question brings up a whole new approach to the idea of regularization. Indeed, if it can be answered positively then it will be possible to analyze power-law and logarithmic divergences with independent scales. In an attempt to find an answer, one comes to realize that an  efficient method is to start from a known regularization scheme and deform it judiciously to arrive at the  ``detached regularization" implied in the question. In this regard,  we repose the question above in a more specialized form
\begin{eqnarray}\label{question-new}
&&How\ to\ extend\ the\ dimensional\ regularization\ scheme\ to\  QFTs\ with\ an\ \ UV\ cutoff\\
&&such\ that\ all\ the\ features\ of\ the\ dimensional\ regularization\ are\ preserved?\nonumber
\end{eqnarray}
In this form, we  construct the detached regularization as an extension or deformation of the dimensional regularization. This new question is actually highly nontrivial because dimensional regularization  scheme is specific to QFTs without an  UV scale \cite{dim-reg1,dim-reg2,dim-reg3,ccb1}. Indeed, the question of how to inject a  cutoff scale in a dimensionally-regularized amplitude has no clear answer. To this end, following the nascent ideas in \cite{demir-talk1,demir-talk2}, in Sec. \ref{sec:det_reg} below, we study the question (\ref{question-new}) by extending the dimensional regularization to $D=0$ and $D=2$ momentum space dimensions  \cite{veltman2,jj1,jj2,jj3,kaplan} at a scale involving not only the renormalization scale $\mu$ but also the  UV cutoff $\Lambda_\wp$ of the QFT. By a judicious structuring, the end result will have $\Lambda_\wp$ setting the power-law UV-sensitivities (replacing $\Lambda$ in $\Lambda^2$ and $\Lambda^4$ terms of cutoff regularization) and $\mu$ parametrizing the logarithmic UV-sensitivities (replacing $\Lambda$ in $\log\Lambda$ terms in cutoff regularization). This two-scale regularization scheme will lead to the sought-for detached regularization if the poles at $D=0$, $D=2$ and $D=4$ are all included. Resting on the dimensional regularization, the detached regularization generates no evanescent contributions different from the ones expected in dimensional regularization at higher loops.    

Having established the detached regularization in Sec. \ref{sec:det_reg},  we apply it to computation of the vacuum energy in a general QFT in Sec. \ref{sec:vac_en}. 

 In Sec. \ref{sec:apps}, we give applications of the detached regularization to two well-known QFTs: the scalar electrodynamics in Sec. \ref{subsec:apps_sqed} and the spinor electrodynamics in Sec. \ref{subsec:apps_qed}. Their calculational details are given in Sec. \ref{Appendices} (Appendices A and B).

In Sec. \ref{sec:subtract}, we illustrate how subtractive renormalization can be realized in the framework of detached regularization. We in particular show that what remains after the subtraction of power-law corrections is just the dimensionally-regularized ${\overline{MS}}$-renormalized QFT.

In Sec. \ref{sec:conc} we conclude.

\section{Detached Regularization}\label{sec:det_reg}
In this section our goal is to answer the question (\ref{question-new}). We want to construct a regularization scheme in which these two properties hold:  
\begin{enumerate}[(a)]
    \item Power-law and logarithmic divergences are both contained in the regularization, and
    \item Power-law and logarithmic divergences involve independent  scales.
\end{enumerate}
To achieve the property (a) we adopt the dimensional regularization scheme \cite{dim-reg1,dim-reg2,dim-reg3} and exploit the fact that dimensional regularization  starts involving quartic and quadratic divergences when the momentum space dimension is set to $D=0$ and $D=2$, respectively \cite{jj1,jj2,jj3,kaplan}. This dimensional change is a highly useful property but it is far from sufficient for achieving the property (b). To that end, we introduce a generalization of the dimensional regularization by  introducing a new scale  $\Lambda_\wp$ besides the usual renormalization scale $\mu$. In explicit terms, we consider an extension of the  form
\begin{eqnarray}
   \mu^{4-D} \int \frac{d^D p}{(2\pi)^D} \frac{1}{(p^2-m^2+i0)^n} \longrightarrow f(\Lambda_\wp, \mu,D) \int \frac{d^D p}{(2\pi)^D} \frac{1}{(p^2-m^2+i0)^n}
   \label{extend}
\end{eqnarray}
in which the new function $f(\Lambda_\wp, \mu,D)$ is to be structured judiciously.  To this end,  we impose the following conditions: 
\begin{enumerate}[(i)]
    \item It should suffice to take $f(\Lambda_\wp, \mu,D)$ as a polynomial  of the form $f(\Lambda_\wp, \mu,D) = \Lambda_\wp^a \mu^b$, where $a$ and $b$ are functions of $n$ and $D$.
    
\item It should be possible to split $f(\Lambda_\wp, \mu,D)$ as $f_{0,2}(\Lambda_\wp, \mu,D)+f_4(\Lambda_\wp, \mu,D)$ such that $f_{0,2}(\Lambda_\wp, \mu,D) \rightarrow 0$ as $D\rightarrow 4$, and  $f_{4}(\Lambda_\wp, \mu,D) \rightarrow 0$ as $D\rightarrow 0,2$.

\item The function $f_{4}(\Lambda_\wp, \mu,D)$ should give the usual dimensional regularization amplitude in the left-hand side of (\ref{extend}).

\item The function $f_{0,2}(\Lambda_\wp, \mu,D)$ should lead to the same powers of $\Lambda_\wp$ compared to the cutoff regularization results in (\ref{int-4d-results}).
\end{enumerate}
 These requirements and limit values  put the regularization function $f(\Lambda_\wp, \mu,D)$ in this compact form
\begin{eqnarray}
f(\Lambda_\wp, \mu,D) = \frac{1}{(8\pi)^{2-n}} \left(\delta_{[D]0} + \delta_{[D]2}\right) \Lambda_\wp^{4-2n} \mu^{2n-D} + \delta_{[D]4} \; \mu^{4-D}
\label{function-f}
\end{eqnarray}
in which $[D]$ designates the integer part of $D$ so that $[0-\epsilon]=0, [2-\epsilon]=2$ and $[4-\epsilon]=4$ for an infinitesimal $\epsilon$. Needless to say, $\delta_{ij}$ is Kronecker delta, which is equal to 1 (0) if $i=j$ ($i\neq j$).  
The normalization factor $1/(8\pi)^{2-n}$ is attached to make  coefficients of the $\Lambda_\wp^{4}$ and $\Lambda_\wp^{2}$ to remain parallel, respectively, to those of the $\Lambda^4$ and $\Lambda^2$ in the cutoff regularization integrals in (\ref{int-4d-results}). In fact, after using $f(\Lambda_\wp, \mu,D)$, the original loop integral in (\ref{int-4d}) takes the form
\begin{eqnarray}
I_{n,D}(\Lambda_\wp,\mu) &=&\Bigg[\frac{1}{(8\pi)^{2-n}} \left(\delta_{[D]0} + \delta_{[D]2}\right) \Lambda_\wp^{4-2n} \mu^{2n-D} + \delta_{[D]4} \; \mu^{4-D}\Bigg]  \int \frac{d^D p}{(2\pi)^D} \frac{1}{(p^2-m^2+i0)^n} \label{int-Dd}\\
&=& \frac{i (-1)^n}{(4 \pi)^{D/2}} \frac{1}{(8\pi)^{2-n}} \frac{\Gamma(n-D/2)}{\Gamma(n)} \left(\delta_{[D]0} + \delta_{[D]2}\right) \Lambda_\wp^{4-2n} \left(\frac{\mu}{m}\right)^{2n-D} \label{int-Dd-2}\\
&+& \frac{i (-1)^n}{(4 \pi)^{D/2}} \frac{\Gamma(n-D/2)}{\Gamma(n)} \delta_{[D]4} \; \mu^{4-2n} \left(\frac{\mu}{m}\right)^{2n-D} \label{int-Dd-3}
\end{eqnarray}
which is seen to be power-law in $\Lambda_\wp$ and yet logarithmic in $\mu$. It is so because  power of $\Lambda_\wp$ is independent of $D$ but that of $\mu$ depends on $D$ and gives rise to  $\log(\mu/m)$ terms when the gamma functions are expanded about momentum space dimensions $D=0,2,4$. (In view of the earlier literature \cite{jj1,jj2,jj3} on power-law divergences in the dimensional regularization scheme, we do not consider odd  dimensions $D=1,3$.) It is clear that the sought-for detachment is achieved: While $\Lambda_\wp$ arises only in power-law terms  $\mu$ appears only in logarithmic terms, and hence, the power-law and logarithmic UV-sensitivities get completely detached. This is the sought-for {\it detached regularization}. The detachment can be explicitly seen by evaluating $I_{n,D}(\Lambda_\wp,\mu)$ for the relevant values of $D$ and $n\leq D/2$
\begin{eqnarray}
I_{n,D}(\Lambda_\wp,\mu) = \left\{\begin{array}{ll} \frac{i}{32 \pi^2} \Lambda_\wp^4 & \quad n=0, D=0\\ \\
 - \frac{i}{32 \pi^2} \Lambda_\wp^2\log \frac{\mu^2}{m^2} & \quad n=1, D=2\\ \\
\frac{i m^2}{16 \pi^2} \left(1 + \log \frac{\mu^2}{m^2}\right) & \quad n=1, D=4\\ \\
\frac{i}{16 \pi^2} \log \frac{\mu^2}{m^2} & \quad n=2, D=4
\end{array}\right.
\label{int-Dd-results}
\end{eqnarray}
after employing the $\overline{MS}$ subtraction scheme  \cite{ccb1,dim-reg3}. These individual loop integrals shed enough light on the roles of the scales $\Lambda_\wp$ and $\mu$. The role of $\mu$ is as usual in that the QFT under consideration runs from scale to scale via the renormalization group equations in terms of $\mu$ \cite{ccb1,dim-reg3}. The role of $\Lambda_\wp$, on the other hand, is also as usual in that it acts as the UV cutoff, as can be seen by comparing (\ref{int-Dd-results}) with  the cutoff regularization integrals in (\ref{int-4d-results}). This is seen also from the fact that  $\Lambda_\wp$ terms arise only in $D=0$  and $D=2$ limits, which correspond, respectively, to the quartic and quadratic UV divergences \cite{jj1,jj2,jj3}. 
The $\Lambda_\wp^4$ term from $D=0$ integral and $\Lambda_\wp^2$ term from the $D=2$ integral both vanish identically in dimensional regularization. Incorporation of these terms by the new regularization method in (\ref{int-Dd}) enables us to take into account the power-law and logarithmic divergences all at once in a way detached from each other.
It is clear that the $\Lambda^{4-2n}_\wp$ factor in (\ref{int-Dd}) is much more than a simple multiplicative factor in that it reveals the UV sensitivity of the QFT as a function of the propagator order $n$. The detachment of the power-law and logarithmic divergences, which was attempted also by the loop regularization method \cite{lr1,lr2,lr3} and by other methods based on implicit regularization  \cite{Cherchiglia},  enables us to analyze the two types of divergences separately and independently.

It is clear that for a proper analysis of the UV behavior of the QFT it is necessary to include each and every pole in (\ref{int-Dd-results}) \cite{ghp1,an-cont-1,kaplan}. Thus, we gather residues of the poles at $D=0,2,4$ to construct the actual loop amplitude in (\ref{int-4d}) 
\begin{eqnarray}
I_n \xrightarrow{\rm detached\ regularization} I_{n}(\Lambda_\wp,\mu) =\sum_{\substack{D=0,2,4 \\ (n\leq D/2)}}
I_{n,D}(\Lambda_\wp,\mu)\ \equiv\ {\rm the\ answer\ to\ the\ question\ in\ eq.\, (\ref{question-new})}
\label{In-result}
\end{eqnarray}
where one keeps in mind that in actual calculations the mass parameter $m^2$ is  a combination of the masses, external momenta, and appropriate Feynman parameters \cite{ccb1,dim-reg3}.  

 Having revealed its effects by the loop integrals in (\ref{int-Dd-results}), it is timely to examine the question of if the detached regularization is unique or not. It actually is unique. In exact terms, if $\Lambda_\wp$ is to remain polynomial (no logarithm of $\Lambda_\wp$) and $\mu$ logarithmic (no polynomial in $\mu$) then the detached regularization is unique to the extent the dimensional regularization itelf is unique. It is so because the power-law divergences in the detached regularization result from the fact that the dimensional regularization generate power-law terms in dimensions $D=0$ and $2$ \cite{jj1,jj2}. In this sense, what is novel in detached regularization is the existence of two separate scales $\Lambda_\wp$ and $\mu$, with distinct roles. To sum up, according to all four requirements below equation (\ref{extend}), the detached regularization prescription in (\ref{function-f}) stands out as a unique prescription.

The cutoff regularization results are given in (\ref{int-4d-results}) above (in $D=4$ momentum space). It is seen that cutoff regularization method leads to power-law divergences $\Lambda^4$ and $\Lambda^2$ as well as the logarithmic divergences $\log(\Lambda^2/M^2)$. Both divergences involve one and the same scale $\Lambda$. Now, let us compare these divergences with the detached regularization results in (\ref{int-Dd-results}). The power-law divergences $\Lambda_\wp^4$ and $\Lambda_\wp^2$ still arise. They are parallel to those in the cutoff regularization results  (excepting the appearance of $\log\mu$ in front of $\Lambda_\wp^2$). The main difference is that logarithmic divergences involve the subtraction scale $\mu$ not the hard momentum cutoff $\Lambda_\wp$ (corresponding to the cutoff $\Lambda$ in equation (\ref{int-4d-results})). In the last two lines of (\ref{int-Dd-results}) there is no involvement of $\Lambda_\wp$ simply because these integrals are for $D=4$. (This is consistent with the structure of dimensional regularization in which power-law divergences are probed by going to $D=0$ and $D=2$ dimensions, as discussed in Refs. [16] and [17]).

To conclude, the detached regularization prescription in equation (\ref{In-result}) constitutes an affirmative answer to the question in (\ref{question-new}). In our derivations we have focused on a typical loop integral like (\ref{int-4d}) but the detached regularization is general enough to apply all loop amplitudes. In fact, we will illustrate this generality when we apply the detached regularization to the vacuum energy in Sec. \ref{sec:vac_en},  the scalar electrodynamics in Sec. \ref{subsec:apps_sqed}, and spinor electrodynamics in Sec. \ref{subsec:apps_qed}. These two QFTs will suffice for illustrating all the important aspects of the detached regularization method. (We will defer the realistic case of the standard model of elementary particles to  future work since it involves  multi-faceted calculations and analyses in the gauge, Higgs and the fermion sectors.)

\section{One-Loop Corrections to Vacuum Energy}\label{sec:vac_en}
In general, one-loop corrections can be cast as a change $\delta S$ in the QFT action $S$ \cite{dim-reg3,ccb1}. In fact, $\delta S$ is a sum over the individual corrections
\begin{eqnarray}
 \delta S_\psi = \frac{i}{2} (-1)^{s_\psi} \int d^4 x \int \frac{d^4 p}{(2\pi)^4} \log \frac{\left(p^2 - m_\psi^2 + T_{int}(\psi,\psi_{other})\right)}{
 M_0^2}  
\end{eqnarray}
for each field $\psi$ of spin $s_\psi$ and mass $m_\psi$. In this formula, $M_0$ is a mass scale and $T_{int}(\psi,\psi_{other})$ collects couplings of $\psi$ to the self and other fields $\psi_{other}$ in the QFT. In the perturbative regime, the $\psi$ action above can be expanded as
\begin{eqnarray}
 \delta S_\psi = \frac{i}{2} (-1)^{s_\psi}  \int d^4 x  \int \frac{d^4 p}{(2\pi)^4}\left[ \underbrace{\log \frac{(p^2 - m_\psi^2)}{M_0^2}}_{\rm non-diagrammatic} + \underbrace{\frac{T_{int}(\psi,\psi_{other})}{p^2-m_\psi^2} -\frac{1}{2} \left(\frac{T_{int}(\psi,\psi_{other})}{p^2-m_\psi^2}\right)^2 + \dots}_{\rm diagrammatic\ (see\ Sec.\ \ref{sec:apps})}\right]
\end{eqnarray}
which splits into a non-diagrammatic part plus a series expansion that can be graphed via Feynman diagrams. The latter will be studied in detail in Sec. \ref{sec:apps} by considering the illustrative cases of the scalar and spinor electrodynamics. The former (non-diagrammatic part) contributes to the vacuum energy, and in the large loop momentum regime it can be recast as 
\begin{eqnarray}
  \int \frac{d^4 {\tilde{p}}}{(2\pi)^4}  - \int \frac{d ^4 p}{(2\pi)^4} \frac{m_\psi^2}{
  p^2-\mu_{IR}^2} - \frac{1}{2} \int \frac{d ^4 p}{(2\pi)^4} \frac{m_\psi^4}{
  (p^2-\mu_{IR}^2)^2}
  \label{recast}
  \end{eqnarray}
after defining $d^4 {\tilde{p}}\equiv d^4 p\log \frac{p^2}{M_0^2}$, introducing an infrared regulator $\mu_{IR}$ which we identify with $m_\psi$ ($\mu_{IR}=m_\psi$), and discarding the terms finite in the UV. Now, using the detached regularization integrals in (\ref{int-Dd-results}) we get from (\ref{recast}) the following $\psi$-contribution to vacuum action 
\begin{eqnarray}
(\delta S_\psi)^{\rm (vac)}= \int d^4 x  \left\{-\frac{(-1)^{s_\psi}}{64 \pi^2} \Lambda_{\wp}^4 +\frac{(-1)^{s_\psi}}{64\pi^2} \Lambda_\wp^2 m_\psi^2 \log \frac{m_\psi^2}{\mu^2} 
+ \frac{(-1)^{s_\psi}}{32 \pi^2}  m_\psi^4 \left(1-\frac{3}{2}\log \frac{m_\psi^2}{\mu^2}\right)  
\right\}
  \label{psi-vacuum}
  \end{eqnarray}
such that inclusion of the contributions of all the  QFT fields leads to the total vacuum action 
\begin{eqnarray}
(\delta S)^{\rm (vac)}= \int d^4 x  \left\{-\frac{(n_b-n_f)}{64 \pi^2}\Lambda_{\wp}^4 +\frac{1}{64\pi^2} \Lambda_\wp^2 {\rm str}\left[M^2 \log \frac{M^2}{\mu^2}\right]  +\frac{1}{32\pi^2}  {\rm str} \left[M^4 \left(1-\frac{3}{2}\log \frac{M^2}{\mu^2}\right)\right]\right\}
  \label{all-vacuum}
  \end{eqnarray}
in which $n_b (n_f)$ is the total number of bosons (fermions) in the QFT, $M^2$ is the mass-squared matrix of fields,  and ${\rm str}[\dots]=\sum_s (-1)^s {\rm tr}[\dots]$ is the supertrace over spins. This vacuum action will be added to the results of the 
 diagrammatic calculations in Sec. \ref{sec:apps} below
(power-law in $\Lambda_\wp$ and logarithmic in $\mu$).  
 
 In the symmetric phase of the QFT where all gauge symmetries are exact (to be spontaneously broken after the loop corrections are included), it turns out that the mass matrix $M$ in (\ref{all-vacuum}) can pertain only to the  {\it scalar} fields, {\it singlet} fermions and {\it vector-like} fermions. It is worth noting that while the quartic term involves all the fields in the QFT via $n_b-n_f$, the quadratic term (in the symmetric phase of the QFT) involves only the scalars and singlet/vector-like fermions. In the SM, only the Higgs field contributes (as four massive physical scalar fields in the symmetric phase). Inclusion of the neutrino masses brings in the right-handed neutrinos as the singlet-fermion sector. Inclusions of dark matter, inflaton, axion and others bring in scalars or singlet/vector-like fermions so that the quadratic correction in (\ref{all-vacuum}) proves to be a sensitive probe of the new particles beyond the SM. These new particles do not have to couple to the SM particles unless required by empirical facts (like inflaton decay and neutrino Majorana masses) or by symmetry reasons (like gauge symmetry and broken supersymmetry).

\section{Applications of the Detached Regularization}\label{sec:apps}
In this section, we shall give applications of the detached regularization to specific loop amplitudes. To this end, we shall perform a detailed analysis of the scalar electrodynamics and spinor electrodynamics as two comprehensive applications of the detached regularization.

\subsection{Scalar Electrodynamics}\label{subsec:apps_sqed}
\begin{figure}[ht!]
\includegraphics[scale=0.95]{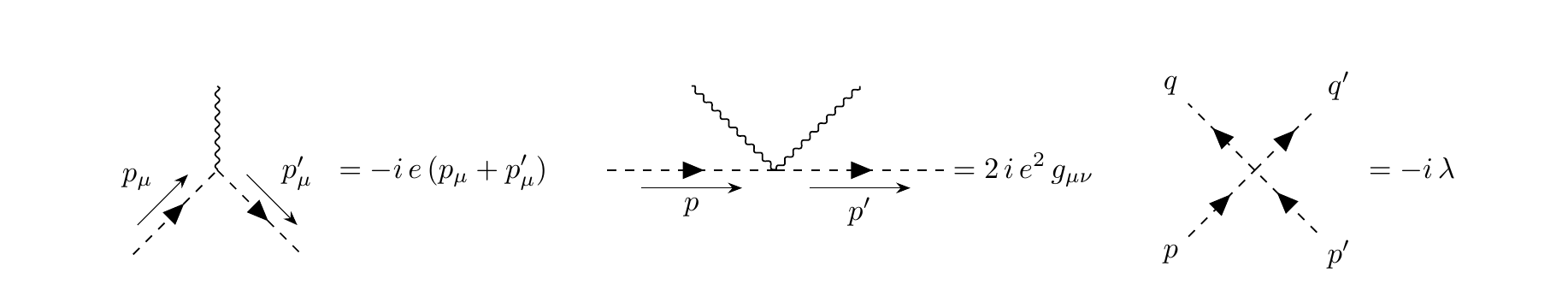}
\caption{The three basic vertices of scalar electrodynamics and their vertex factors.}
\label{fig:sqed_ver}
\end{figure}

In order to illustrate the use of detached regularization, we consider in this subsection a simple QFT composed of a charged spin-zero particle ($\phi$) and the  gauge field ($A_\mu$) --  the so-called \textit{scalar electrodynamics}. The Lagrangian defining this theory is
\begin{equation}\label{sqed_lag}
    {\cal L}=(D^{\mu}\phi)^\dag (D_{\mu}\phi)-m^2 \phi^{\dag} \phi -\frac{1}{4}\lambda (\phi^{\dag} \phi)^2 -\frac{1}{4} F^{\mu \nu}F_{\mu \nu}
\end{equation}
where $F_{\mu \nu}$ is the field strength tensor and $D_{\mu}$ is the gauge covariant derivative defined as
\begin{equation}
    D_{\mu}=\partial_\mu -i e A_\mu \, .
\end{equation}
The Lagrangian in (\ref{sqed_lag}) leads to the three interaction vertices depicted in FIG. \ref{fig:sqed_ver}.

\begin{figure}[ht!]
\begin{tabular}{cc}
  \includegraphics[width=65mm]{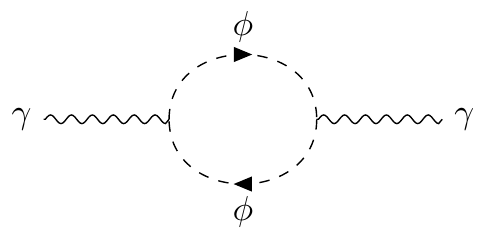} &   \includegraphics[width=65mm]{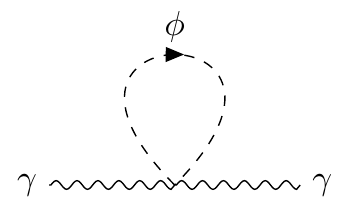} \\
(a)  & (b)  \\[6pt]
\end{tabular}
\caption{One-loop corrections to the photon propagator in scalar electrodynamics.}
\label{fig:sqed_photon}
\end{figure}

Now, we will illustrate the usage of the newly-developed detached regularization in the context of loop corrections in scalar electrodynamics. The details of the calculation can be found in Appendix A at the end of the paper. The relevant diagrams fall into two main categories. The first category concerns the one-loop corrections to the photon propagator due to the charged scalar which is given in FIG. \ref{fig:sqed_photon}.

It is more convenient to combine the self-energy and tadpole diagrams given in FIG. \ref{fig:sqed_photon} (a) and \ref{fig:sqed_photon} (b) into a single amplitude ($k_\mu$ ($p_\mu$) is loop (external) momentum)  
\begin{equation}
    i\Pi^{\mu\nu}_{\ref{fig:sqed_photon}a+\ref{fig:sqed_photon}b}\,(p)=e^2\int\frac{d^4 k}{(2\pi)^4}\, \frac{(2k+p)^{\mu}(2k+p)^{\nu}-2g^{\mu\nu}[(k+p)^2-m^2]}{[(k+p)^2-m^2][k^2-m^2]} \, ,
\end{equation}
from which it will be easier to see the eventual transversality of the logarithmic part. Indeed,  applying the detached regularization scheme in (\ref{int-Dd}) to this amplitude, we get in the $\overline{MS}$ subtraction scheme (see Appendix A)

\def\at{
  \left.
  \vphantom{\int}
  \right|
}

\begin{equation}
    i\Pi^{\mu\nu}_{\ref{fig:sqed_photon}a+\ref{fig:sqed_photon}b}\,(p)= -\:\frac{ i e^2 }{16\pi^2}\, g^{\mu \nu}\; \Lambda_\wp^2 \: -\frac{ i e^2 }{4\pi^2}\: [p^2g^{\mu\nu}-p^{\mu} p^{\nu}]\; \int\limits_{-1/2}^{1/2} dy\Bigg(y^2\: \log \Bigg[\frac{\mu^2}{p^2(y^2-\frac{1}{4})+m^2}\Bigg]\Bigg) \; ,
    \label{eqn:photon_prop_sqed}
\end{equation}
in which the first term is the power-law part (similar to what one would find by employing the cutoff regularization results in (\ref{int-4d-results})), and the second term  which is the finite-transverse part we are familiar from the dimensional regularization \cite{ccb1,dim-reg3}.

\begin{figure}[h!]
\begin{tabular}{ccc}
  \includegraphics[width=55mm]{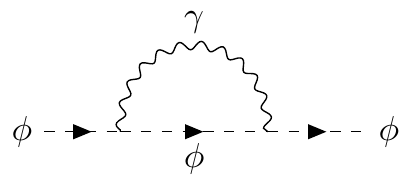} &   \includegraphics[width=55mm]{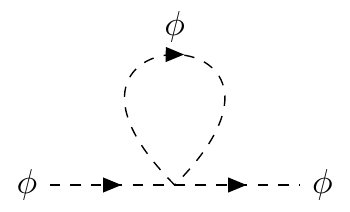} & \includegraphics[width=55mm]{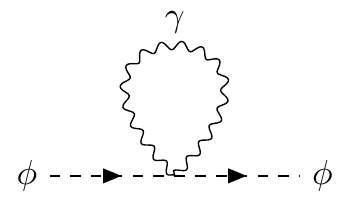} \\
(a)  & (b) & (c) \\[6pt]
\end{tabular}
\caption{One-loop corrections to the scalar propagator in scalar electrodynamics.}
\label{fig:sqed_scalar}
\end{figure}

The second class of loop corrections that is going to be examined in scalar electrodynamics is the one-loop corrections to the scalar propagator itself. The relevant diagrams pertaining to this type of correction is given in FIG. \ref{fig:sqed_scalar}. The self-energy correction in FIG. \ref{fig:sqed_scalar} (a) is given by the amplitude (see Appendix A)

\begin{equation}
    i \Pi_{\ref{fig:sqed_scalar}a}(p)= -e^2\int\frac{d^4 k}{(2\pi)^4}\, \frac{P_{\mu\nu}(k) (k+2p)^{\mu} (k+2p)^{\nu}}{k^2 \, [(k+p)^2-m^2]}
\end{equation}
in which
\begin{equation}\label{eqn:photon_prop}
    P_{\mu\nu}(k)= g_{\mu\nu}-\frac{k_{\mu}k_{\nu}}{k^2}
\end{equation}
is the projector of the photon propagator in Lorenz gauge. The detached regularization scheme leads to 

\begin{equation}
    i \Pi_{\ref{fig:sqed_scalar}a}(p)= \frac{-3\, i\, e^2 \, p^2}{16\pi^2}\, \Bigg\{\log \;  \Bigg(\frac{\mu^2}{m^2-p^2}\Bigg) - \frac{m^2}{p^2}\, \log \Bigg(\frac{m^2}{m^2-p^2}\Bigg) + \frac{4}{3}\Bigg\} \; ,
\end{equation}
where  $\overline{MS}$ subtraction scheme is applied again. This result is essentially the same as the one you would obtain in the dimensional regularization framework \cite{ccb1,dim-reg3}. 


The scalar-scalar tadpole correction in FIG. \ref{fig:sqed_scalar} (b) is given by the amplitude

\begin{equation}
    i \Pi_{\ref{fig:sqed_scalar}b}= \lambda \int\frac{d^4 k}{(2\pi)^4}\, \frac{1}{k^2 -m^2}
\end{equation}
where $m$  is the mass of the charged scalar. The detached regularization (\ref{int-Dd}) with the $\overline{MS}$ subtraction results in the amplitude (see Appendix A) 

\begin{equation}
    i \Pi_{\ref{fig:sqed_scalar}b}= \frac{- i\, \lambda}{32\pi^2}\, \Lambda_\wp^2 \, \log \;  \Bigg(\frac{\mu^2}{m^2}\Bigg) \; +\frac{ i\, \lambda}{16\pi^2}\, m^2 \, \Bigg\{\log \;  \Bigg(\frac{\mu^2}{m^2}\Bigg) +1\Bigg\} \;.
    \label{eqn:phi4_2p}
\end{equation}
It is interesting to note that the second term is the same as the one you would get in dimensional regularization scheme \cite{ccb1,dim-reg3}. The first term, on the other hand,  is rather specific to the detached regularization method (\ref{int-Dd}).

The third and last diagram to consider in scalar electrodynamics is the one in FIG. \ref{fig:sqed_scalar} (c), which is the scalar-photon tadpole diagram. The amplitude for this is

\begin{equation}
    i \Pi_{\ref{fig:sqed_scalar}c}= 2\; e^2\; \int\frac{d^4 k}{(2\pi)^4}\, \frac{g^{\mu \nu} P_{\mu \nu}(k)}{k^2 -m_{\gamma}^{2}}
\end{equation}
where $P_{\mu \nu}(k)$ is again given by (\ref{eqn:photon_prop}). Here, we added a fictitious photon mass $m_{\gamma}$ to the photon propagator as an infrared regulator. The detached regularization scheme (\ref{int-Dd}) applied to this diagram results in the following amplitude after the $\overline{MS}$ subtraction (see Appendix A):

\begin{equation}
    i \Pi_{\ref{fig:sqed_scalar}c}= \frac{ i\, e^2}{8\pi^2}\, \Lambda_\wp^2 \; - \frac{ i\, e^2}{16\pi^2}\, \Lambda_\wp^2 \log \Bigg( \frac{\mu^2}{m_{\gamma}^{2}}\Bigg) 
\end{equation}
which would vanish in the dimensional regularization scheme (more precisely, the first term would disappear and $\Lambda_\wp^2$ in the second term would be replaced by $m_\gamma^2$). This result, which is quadratic in the  scale $\Lambda_\wp$ and logarithmic in the renormalization scale $\mu$, is specific to the detached regularization scheme  (\ref{int-Dd}).

Before closing, it proves useful to dwell on the loop-induced photon mass in equation (\ref{eqn:photon_prop_sqed}). In this regard, one notes that different regularization schemes can be contrasted by typical scattering processes such as the Drell-
Yan scattering $e^+ e^- \rightarrow \gamma^\star \rightarrow f {\bar{f}}$ ($f=$ leptons, quarks) (see \cite{regularization} for a detailed study). The detached regularization has a different take compared to those in \cite{regularization} in that it aims at revealing first the effects of the UV cutoff. Indeed, it is clear that photon acquires a mass (as in (\ref{eqn:photon_prop_sqed}) above) in the presence of the cutoff $\Lambda_\wp$, and consequently, the Drell-Yan cross section exhibits resonance behavior at the loop-induced photon mass. But it is also clear that a finite photon mass is unphysical despite the fact that the UV cutoff exists as a concrete scale. This means that one has to do something about the UV cutoff.  That ``something" could be Sakharov's induced gravity \cite{sakharov1,sakharov2} ($\Lambda_\wp$ leads to Newton's constant) or Demir's emergent gravity \cite{demir1,demir2,demir3} ($\Lambda_\wp$ is promoted to affine curvature as a spurion in resemblance to the Higgs mechanism) or the subtractive renormalization scheme ($\Lambda_\wp^2$ terms are subtracted away) or some other mechanism. Our goal in the present work is to prepare a framework (like the photon polarization as in (\ref{eqn:photon_prop_sqed}) above) in which one can do ``something" about the UV cutoff such that what is left after that ``something" is the usual dimensionally-regularized QFT. In other words, after the UV cutoff $\Lambda_\wp$ is dealt with the Drell-Yann scattering $e^+ e^- \rightarrow \gamma^\star \rightarrow f {\bar{f}}$ proceeds as in the dimensional regularization at the subtraction scale $\mu$ (as discussed in (\ref{eqn:photon_prop_sqed}), for instance).

\subsection{Spinor Electrodynamics}\label{subsec:apps_qed}
We continue applications of the detached regularization with the {\it spinor electrodynamics}, which is a simple QFT composed of a charged Dirac fermion ($\psi$) and the gauge field ($A_\mu$). The calculational details of this section are all given in Appendix B. The theory is governed by the Lagrangian 
\begin{equation}
    {\cal L}= -\frac{1}{4} F^{\mu \nu}F_{\mu \nu} + \overline{\psi} \left(i\slashed{\partial} - e \slashed{A} -m_{f} \right) \psi
\end{equation} 
where $F_{\mu \nu}$ is the field strength tensor for $A_\mu$.
\begin{figure}[ht!]
\begin{tabular}{ccc}
  \includegraphics[width=55mm]{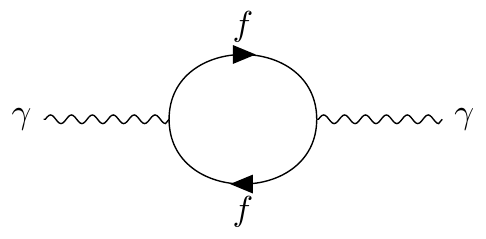} &   \includegraphics[width=55mm]{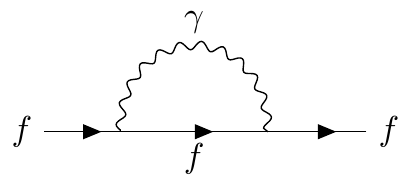} & \includegraphics[width=55mm]{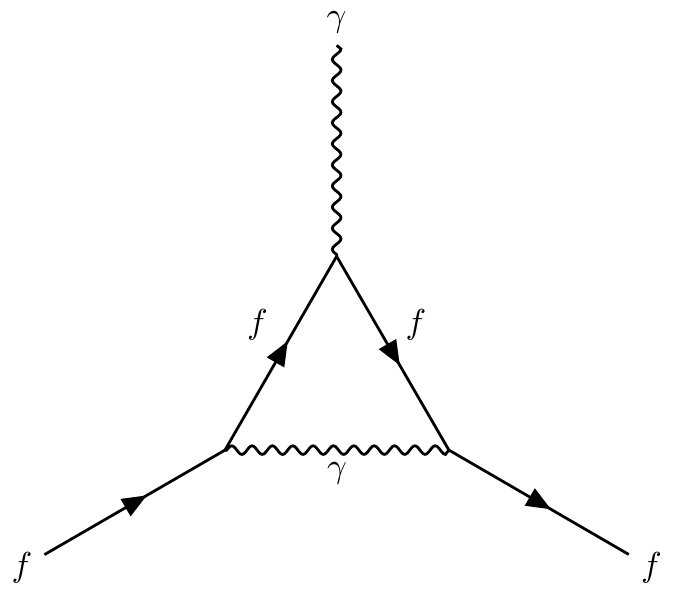} \\
(a)  & (b) & (c) \\[6pt]
\end{tabular}
\caption{One-loop corrections in spinor electrodynamics.}
\label{fig:qed}
\end{figure}
In regard to the applications of the detached regularization method  in (\ref{int-Dd}), we will calculate the amplitudes for the one-loop Feynman diagrams  depicted in FIG. \ref{fig:qed}. 

We begin with the photon vacuum polarization diagram in FIG. \ref{fig:qed} (a). Its  is given by

\begin{equation}
i\Pi^{\mu \nu}_{\ref{fig:qed}a}(p)=-e^2\int \frac{d^4 k}{(2\pi)^4}\frac{{\rm Tr}\left[\gamma^{\mu}(\slashed{k}+\slashed{p}+m_{f})\gamma^{\nu} (\slashed{k}+m_{f})\right]}{\left[k^2-m_{f}^{2}\right]\left[(k+p)^2-m_{f}^{2}\right]} \; .
\end{equation}
Regularization of  this amplitude by the detached regularization in (\ref{int-Dd}) leads to (see Appendix B)

\begin{equation}
i\Pi^{\mu \nu}_{\ref{fig:qed}a}(p)=\frac{ i e^2 }{16\pi^2}\, g^{\mu \nu}\; \Lambda_\wp^2 \: -\frac{ i e^2 }{2\pi^2}\: [p^2g^{\mu\nu}-p^{\mu} p^{\nu}]\; \int\limits_{0}^{1} dx \left\{\: (x-x^2)\:\left(- \:\,\frac{1}{2} + \log \left[\frac{\mu^2}{p^2(x^2-x)+m_{f}^{2}}\right]\right)\right\} \; ,
\label{eqn:photon_prop_qed}
\end{equation}
whose first term is a quadratic correction (similar to what one would get by cutoff regularization). Its second term is finite and transverse just like the corresponding vacuum polarization diagram in (\ref{eqn:photon_prop_sqed}) in the scalar electrodynamics. It is what one  would get from the dimensional regularization \cite{ccb1,dim-reg3}.

The fermion self-energy diagram in FIG. \ref{fig:qed} (b) obtains the amplitude

\begin{equation}
-i\Sigma_{\ref{fig:qed}b}(p)=-e^2\int \frac{d^4 k}{(2\pi)^4}\frac{\gamma^{\mu}(\slashed{k}+m_{f})\gamma_{\mu}}{\Big[k^2-m_{f}^2\Big]\Big[(p-k)^2-m_{\gamma}^2\Big]} \; 
\end{equation}
and its regularization by the detached regularization in (\ref{int-Dd}) results in (see Appendix B)

\begin{equation}
-i\Sigma_{\ref{fig:qed}b}(p)= \frac{ i e^2 }{16\pi^2}\:\int\limits_{0}^{1} dx \left[2\, m_{f} - 2 x\, \slashed{p}\right]\; - \;  \frac{ i e^2 }{16\pi^2}\:\int\limits_{0}^{1} dx \left\{\left[4\, m_{f} - 2 x\, \slashed{p}\right]\;\log \left[\frac{\mu^2}{p^2(x^2-x) + x \, m_{\gamma}^2 + (1-x)\, m_{f}^{2}}\right]\right\} 
\end{equation}
after applying the $\overline{MS}$ subtraction scheme.

Lastly, the fermion-photon vertex of diagram in FIG. \ref{fig:qed} (c) has the amplitude
\begin{equation}
-i e \Gamma^{\mu}_{\ref{fig:qed}c}(p)=-e^3\int \frac{d^4 k}{(2\pi)^4}\frac{\gamma_{\nu}(\slashed{p}^{'}-\slashed{k}+m_{f})\gamma^{\mu} (\slashed{p}-\slashed{k}+m_{f})\gamma^{\nu}}{k^2 \, \Big[(p-k)^2-m_{f}^2\Big]\Big[(p^{'}-k)^2-m_{f}^2\Big]} \; 
\end{equation}
whose detached regularization (\ref{int-Dd}) gives rise to (see Appendix B)

\begin{equation}
-i e \Gamma^{\mu}_{\ref{fig:qed}c}(p)=-2\,e^3 \int\limits_{0}^{1} dx \, \int\limits_{0}^{1-x} dy  \left\{\frac{ i \gamma^{\mu} }{16\pi^2}\left(-\frac{5}{2}\,+\,\log \frac{\mu^2}{\Delta^{2}}\right)\,-\, \frac{ i \tilde{N}^{\mu} }{32\,\pi^2\, \Delta^{2}}\right\}
\end{equation} 
in which
\begin{equation}
    \Delta^{2}=\, m_{f}^2 \, (x+y) + p^2 \, (x^{2}-x) + {p^{'}}^{2} \, (y^{2}-y)\,+2\,p \cdot p^{'} (xy) 
\end{equation}
and 
\begin{equation}
    \tilde{N}^{\mu}= \gamma_{\nu}\left[\slashed{p}^{'}(1-y)-\slashed{p}x+m_{f}\right]\gamma^{\mu} \left[\slashed{p}(1-x)-\slashed{p}^{'}y+m_{f}\right]\gamma^{\nu} \; .
\end{equation}
Needless to say, the fermion-photon vertex is what one would find in dimensional regularization. It is an important property of the detached regularization that all the results and properties of the dimensional regularization are maintained. 

In parallel with the discussion at the end of Sec. IV A, here it should be emphasized that the goal in the present work is to construct a framework (detached regularization scheme) in which one can deactivate the photon mass in equation (\ref{eqn:photon_prop_qed}) in a way leaving behind only a dimensionally-regularized QFT.  It is after the deactivation of the photon mass that the Drell-
Yan scattering $e^+ e^- \rightarrow \gamma^\star \rightarrow f {\bar{f}}$ ($f=$ leptons, quarks) proceeds as in the dimensional regularization \cite{regularization}. In the next section, we discuss how the subtractive renormalization can be naturally realized in the framework of detached regularization.

\section{Subtractive renormalization in the framework of detached regularization } \label{sec:subtract}

As was mentioned in Introduction, subtractive renormalization \cite{ quadratic-logarithmic2,quadratic-logarithmic1} is one instance in which distinction between power-law and logarithmic divergences is a must. The detached regularization constructed in Sec. II is one such distinctive regularization framework. Its applications  in Sec. III and Sec. IV (see also Appendix A and Appendix B for details) have shown that the power-law and logarithmic divergences get manifestly detached from each other under the regularization function $f(\Lambda_\wp, \mu,D)$ in (\ref{function-f}) (supplemented with the ${\overline{MS}}$ subtraction). 

One immediate field theoretic setup in which  one can  benefit from the merits of the detached regularization is the subtractive renormalization of the $\lambda \phi^4$ theory. To this end, we analyze the scalar two-point function, and show how subtractive renormalization works in the detached regularization scheme.  One recalls that in subtractive renormalization \cite{ quadratic-logarithmic2,quadratic-logarithmic1}, one subtracts out the quadratically divergent ($\Lambda_\wp^2$)  terms by adding relevant counter terms to the theory. However, since the commonly employed regularization methods do not detach power-law and logarithmic dependencies, mass renormalization factor $Z_{m}$ still contains a term proportional to the logarithm of $\Lambda_\wp^2$ such as \cite{quadratic-logarithmic2} 
\begin{eqnarray}
    \frac{\lambda}{32\pi^{2}}\log\left(\frac{\Lambda_\wp^2}{m^{2}}\right)
\end{eqnarray}
for a real scalar with mass $m$. This means that even though the quadratic terms ($\Lambda_\wp^2$) are subtracted out the remnant logarithmically divergent terms ($\log \Lambda_\wp$) continue to involve the UV cutoff $\Lambda_\wp^2$. 

This remnant $\Lambda_\wp$ problem does not arise in the detached regularization. Indeed, in detached regularization two-point function of a real scalar field $\phi$ follows from (\ref{eqn:phi4_2p}) after scaling by $1/2$, and takes the compact form   
\begin{eqnarray}\label{eqn:two-point}
\Gamma^{(2)}(p^2)= p^2-m^2-\frac{\lambda}{64\pi^2}\Lambda_\wp^2\log\left(\frac{\mu^2}{m^2}\right)+\frac{\lambda}{32\pi^2}m^2+\frac{\lambda}{32\pi^2}m^2\log\left(\frac{\mu^2}{m^2}\right)
\end{eqnarray}
where $p$ is the momentum of the particle. Now, application of the subtractive renormalization \cite{quadratic-logarithmic2}  to this two-point function removes the $\Lambda_\wp^2$ term to leave behind $\log\mu$--involving terms as in the dimensional regularization \cite{ccb1,dim-reg3}. It is in this sense that it becomes possible to subtract away power-law UV-sensitivities such that what remains after the subtraction is precisely what one would find in the dimensional regularization (with no dependence on $\Lambda_\wp$). In the language of \cite{ quadratic-logarithmic1}, the local power-law terms are cleaned of the regularized theory in a way causing no physical effects.

\begin{figure}[h!]
\begin{center}
\includegraphics[scale=1.3]{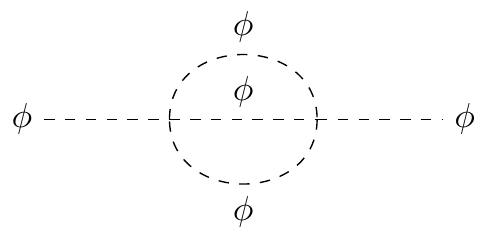}
\end{center}
\caption{Two-loop sunset diagram for a real scalar field $\phi$ with $\lambda \phi^4$ coupling.}
\label{fig:two_loop}
\end{figure}

Having done with the subtractive renormalization of the one-loop self energy in detached regularization, it is now time to discuss workings of the detached regularization at two-loop order. This can be done by analyzing the sunset diagram in Fig. \ref{fig:two_loop}, for instance. The goal is to determine if the regularization function 
$f(\Lambda_\wp, \mu,D)$ in (\ref{function-f}) properly works at two loops. The superficial degree of divergence of the sunset diagram is two. It therefore is expected to be quadratically divergent (proportional to $\Lambda_\wp^2$). In fact, direct calculation gives

\begin{equation}
    -i\Sigma=\frac{i \lambda^2}{96 \pi^2}\int\int\int dx \, dy \, dz \, \delta(x+y+z-1)\frac{1}{\sqrt{\alpha \beta}}\Bigg\{\frac{\Lambda_\wp ^2}{32 \pi^2} \log\frac{\mu^2}{{\overline  m}^2}-\frac{{\overline m}^2}{16\pi^2} \Big(\log\frac{\mu^2}{{\overline m}^2}+1\Big)\Bigg\}
    \label{isigma}
\end{equation}
in which
\begin{equation}
    \alpha= x+z \, , \quad  \beta=\frac{xz+y(x+z)}{x+z} \, , \quad \gamma=\frac{xyz}{xz+y(x+z)}\,, \quad {\overline m}^2=-\gamma p^2+(x+y+z)m^2\,,
\end{equation}
and $m$ is the mass of the scalar, $p$ is the external momentum. The first term inside the curly bracket in (\ref{isigma}) is the aforementioned  quadratic divergence of the sunset diagram in Fig. \ref{fig:two_loop}. The second term in (\ref{isigma}), on the other hand,  is the expression one would find in dimensional regularization. Clearly, the two-loop self energy piece (\ref{isigma})  ensures that, at higher loops, the detached regularization continues to comprise both the power-law ($\Lambda_\wp^2$) and logarithmic ($\log\mu$) UV-sensitivities. As mentioned above while discussing the one-loop self-energy, subtractive renormalization weeds out $\Lambda_\wp^2$ terms to leave behind a $\log\mu$-involving expression. All this implies that the detached regularization provides a natural setting for subtractive renormalization at and beyond one loop.

\section{Conclusion} \label{sec:conc}
 The dimensional regularization sets the common ground for regularizing QFTs (see Ref. \cite{regularization} for various variants) thanks mainly to its ability to preserve 
the gauge invariance.  It gives a gauge-invariant description of how the QFT changes with the renormalization scale $\mu$ (essentially the scale of the experiment). But the dimensional regularization holds good provided that the QFT under concern is devoid of any cutoff scale. (The cutoff $\Lambda_\wp$ is not a particle mass. It is the UV boundary of the QFT in the Wilsonian sense). The cutoff does necessarily break the gauge symmetries (violation of the Ward identities). In the present work, the goal has been to extend the dimensional regularization to QFTs with a UV cutoff such that power-law divergences (involving $\Lambda_\wp$ and breaking gauge symmetry) and logarithmic divergences (involving $\mu$ and respecting gauge symmetry) are combined in one single regularization scheme. This detachment of the two divergences is motivated by their natures. Indeed, power-law divergences are local in that they live at the scale $\Lambda_\wp$. The logarithmic divergences, on the other hand, extend in the entire range since the correction $M^2 \log \Lambda_\wp^2/M^2$ varies with the particle mass $M$. Besides, the logarithmic UV-sensitivities set the beta functions and conformal anomalies in the QFT while the power-law divergences do not play such roles.

In addition to the above, there is the fact that the conventional cut-off regularization does not assign different mass scales to the power-law and logarithmic divergences. In the case of the induced gravity the UV cutoff sets Newton's constant. In the case of emergent gravity, on the other hand, the UV cutoff gets promoted to affine curvature as a spurion field (similar in philosophy to the Higgs mechanism). Finally, in the subtractive renormalization procedure, power-law divergences are cancelled away by the introduction of appropriate counter terms. In all these mechanisms,  the problem boils down to the fact that while one operates on the terms power-law in $\Lambda_\wp$ (whether it be identification or promotion or cancellation)  one has to overlook  the dependence of the logarithmic terms on the same scale $\Lambda_\wp$. We illustrated this problem by a discussion of the subtractive renormalization in Sec. \ref{sec:subtract}. In that case, even though one subtracts away the power-law dependencies on $\Lambda_\wp$, one is still left with the $\log \Lambda_\wp$ dependencies which survive in gauge-invariant corrections (including the fermion masses). It is clear that if one had a regularization method that separates out the scales on which the power-law and logarithmic divergences depend, all these predicaments would evaporate. This is the main motivation of the present manuscript and it is the main supremacy of detached regularization over the cut-off regularization.

 In this work, for the first time in the literature, we have extended the usual dimensional regularization to involve the logarithmic ($D=4$) and power-law UV sensitivities ($D=0$ and $D=2$) in a way detached to involve different scales. We have demonstrated benefits of the detached regularization by giving its basic uses in different circumstances.  Firstly, we have applied it to the computation of the vacuum energy in Sec. \ref{sec:vac_en}, finding the usual UV structure. Secondly, we have applied it to regularization of the scalar and spinor electrodynamics in Sec. \ref{sec:apps}. Thirdly, we have applied it to subtractive renormalization to show how eligible the detached regularization is for treating the power-law and logarithmic divergences independently. In general, the detached regularization is unique in that it separates out the scales of the power-law and logarithmic corrections, where the latter sets the running of the parameters of the theory by virtue of the beta functions.  

The detachment of the power-law and logarithmic divergences can be of broad interest for the renormalization of QFTs. The renormalization schemes in which both logarithmic and power-law divergences are kept can prove useful (as did in nuclear physics applications \cite{kaplan}) for revealing the UV sensitivity of the QFTs. If one wants to renormalize away the quadratic UV dependencies, one way to go by is to employ the subtractive renormalization \cite{quadratic-logarithmic1,quadratic-logarithmic2}. In conventional subtractive renormalization  quadratic and logarithmic divergences formally appear as separate terms but the logarithmic terms  embedded in the bare mass still inhabit the same scale as the power-law ones which are subtracted out via counter terms. This may seem like a simple issue. However it is more involved than that, since the renormalized mass term depends on the scale $\Lambda_\wp$. In contrast to this conventional structure, the newly-introduced detached regularization enables a complete detachment of the power-law and logarithmic divergences, and this detachment in return enables the subtractive renormalization to subtract away all quadratic divergences, leaving behind exactly the logarithmic terms one would find in the dimensional regularization. 

The detached regularization can have potential applications in various problems. It can be utilized in a broad class of renormalization methods or field-theoretic mechanisms.  One example is subtraction of the power-law divergences as in \cite{quadratic-logarithmic1,quadratic-logarithmic2,quadratic-logarithmic3}. Another example is induced gravity models in which UV cutoff leads to Newton's constant \cite{sakharov1,sakharov2}.  Yet another example is emergent gravity models in which UV cutoff is promoted to curvature (in reminiscence to promotion of vector boson masses to Higgs field) \cite{demir1,demir2,demir3}. These examples can be furthered with other possible applications.

\section*{Acknowledgements}

The work of O. S. is supported by the T{\"U}B{\.I}TAK B{\.I}DEB-2218 national postdoctoral fellowship grant 118C522. D. D.  acknowledges the contribution of the COST Action
CA21106 - COSMIC WISPers in the Dark Universe: Theory, astrophysics and experiments (CosmicWISPers). The authors are grateful to conscientious reviewers for their comments, criticisms and suggestions.


\newpage

\section{Appendices} \label{Appendices}
\subsection*{Appendix A: Scalar Electrodynamics} 
\label{AppendixA}
\subsubsection*{Vacuum polarization in Scalar Electrodynamics}
Combining the self-energy and tadpole diagrams given in FIG. \ref{fig:sqed_photon} (a) and \ref{fig:sqed_photon} (b) we obtain
\begin{equation}
    i\Pi^{\mu\nu}_{\ref{fig:sqed_photon}a+\ref{fig:sqed_photon}b}\,(p)=e^2\int\frac{d^4 k}{(2\pi)^4}\, \frac{(2k+p)^{\mu}(2k+p)^{\nu}-2g^{\mu\nu}[(k+p)^2-m^2]}{[(k+p)^2-m^2][k^2-m^2]} \, ,
\end{equation}
to which we apply Feynman parametrization in the denominator and shift the loop momenta in the numerator accordingly. Getting rid of the terms linear in the shifted loop momenta $q$, we get to the point
\begin{eqnarray}\label{eqn:sqed_phot1}
    i\Pi^{\mu\nu}_{\ref{fig:sqed_photon}a+\ref{fig:sqed_photon}b}\,(p)&=&e^2\int\limits_{0}^{1} dx\;\left\{g^{\mu\nu}\left(\frac{4}{D}-2\right)\int\frac{d^4 q}{(2\pi)^4}\,\frac{q^2}{\left[q^{2}-\Delta^{2}\right]^2} \;\right\}  \\ \nonumber
    &+&\;e^2\int\limits_{0}^{1} dx\;\left\{ \left[p^{\mu}p^{\nu}(2x-1)^2-2g^{\mu\nu}[p^2(1-x)^2-m^2]\right]\int\frac{d^4 q}{(2\pi)^4}\,\frac{1}{\left[q^{2}-\Delta^{2}\right]^2} \right\}
\end{eqnarray}
where
\begin{equation} \label{eqn:feyn_delta}
    \Delta^2=p^2(x^2-x)+m^2 \: ,
\end{equation}
and the first integral in (\ref{eqn:sqed_phot1}) can be reduced to two irreducible integrals. The result of this reduction is to take the amplitude (\ref{eqn:sqed_phot1}) into

\begin{equation}
    i\Pi^{\mu\nu}_{\ref{fig:sqed_photon}a+\ref{fig:sqed_photon}b}\,(p)=e^2\int\limits_{0}^{1} dx\;\Big\{I_1+I_2\Big\}
\end{equation}
where
\begin{equation}\label{eqn:i1}
    I_1=g^{\mu\nu}\left(\frac{4}{D}-2\right)\int\frac{d^4 q}{(2\pi)^4}\:\frac{1}{q^{2}-\Delta^{2}}
\end{equation}
and 
\begin{equation}\label{eqn:i2}
    I_2=g^{\mu\nu}\left(\frac{4}{D}-2\right)\, \Delta^2\;\int\frac{d^4 q}{(2\pi)^4}\:\frac{1}{\left[q^{2}-\Delta^{2}\right]^2}\,+\,\left[p^{\mu}p^{\nu}(2x-1)^2-2g^{\mu\nu}[p^2(1-x)^2-m^2]\right]\int\frac{d^4 q}{(2\pi)^4}\,\frac{1}{\left[q^{2}-\Delta^{2}\right]^2} \: .
\end{equation}

Now, we should first apply the new regularization (\ref{int-Dd}) to $I_1$ in (\ref{eqn:i1}). The power  of the denominator of $I_1$ is $n=1$. Therefore (\ref{int-Dd}) applied to (\ref{eqn:i1})  reads
\begin{equation}\label{eqn:i1_1}
    I_1=g^{\mu\nu}\left(\frac{4}{D}-2\right)\;\Bigg[\frac{1}{8\pi}\left(\delta_{[D]0}+\delta_{[D]2}\right) \Lambda_\wp^{2} \: \: \mu^{2-D} + \delta_{[D]4}\; \mu^{4-D}\Bigg] \;\int\frac{d^D q}{(2\pi)^D}\:\frac{1}{q^{2}-\Delta^{2}}
\end{equation}
where the integral on the RHS amounts to
\begin{equation}\label{eqn:int_gamma1}
    \int\frac{d^D q}{(2\pi)^D}\:\frac{1}{q^{2}-\Delta^{2}} = \frac{-i}{(4 \pi)^{D/2}} \frac{\Gamma(1-D/2)}{\Gamma(1)}(\Delta^2)^{D/2-1} \: .
\end{equation}
Replacing (\ref{eqn:int_gamma1}) into (\ref{eqn:i1_1}) we obtain the following expression which is a function of the dimensionality $D$

\begin{equation}\label{eqn:i1_2}
    I_1=-i\; g^{\mu\nu}\left(\frac{4}{D}-2\right)\;\Bigg[ \frac{1}{8 \pi}\left(\delta_{[D]0}+\delta_{[D]2}\right) \Lambda_\wp^{2} \: \: \mu^{2-D} + \delta_{[D]4} \; \mu^{4-D}\Bigg] \;\frac{(\Delta^2)^{D/2-1}}{(4 \pi)^{D/2}} \frac{\Gamma(1-D/2)}{\Gamma(1)} \: .
\end{equation}
Now, the crucial point to keep in mind is that in evaluating (\ref{eqn:i1_2}), not only do we have to consider the $D\to4$ limit but we should also take the $D\to2$ limit so that we don't leave out the quadratic corrections in $\Lambda_\wp$. In the $D\to2$ limit (\ref{eqn:i1_2}) yields
\begin{equation}\label{eqn:i1_3}
    (I_1)_{D\to2}=-i\; g^{\mu\nu} \: \frac{\Lambda_\wp^{2}}{16\pi^2}
\end{equation}
making use of the $\overline{MS}$ subtraction scheme. 

Before calculating  the $D\to4$ limit of $I_1$, first let us evaluate $I_2$ in (\ref{eqn:i2}). The power of the denominator of the divergent integrals in $I_2$ is $n=2$, therefore the new regularization scheme (\ref{int-Dd}) applied to it reads

\begin{eqnarray}\label{eqn:i2_1}
    &I_2&=g^{\mu\nu}\left(\frac{4}{D}-2\right)\, \Delta^2 \Bigg[\left(\delta_{[D]0}+\delta_{[D]2}\right)  \mu^{4-D} + \delta_{[D]4} \; \mu^{4-D}\Bigg]\:\int\frac{d^D q}{(2\pi)^D}\,\frac{1}{\left[q^{2}-\Delta^{2}\right]^2} \\ \nonumber
        &+&\left[p^{\mu}p^{\nu}(2x-1)^2-2g^{\mu\nu}[p^2(1-x)^2-m^2]\right]\Bigg[\left(\delta_{[D]0}+\delta_{[D]2}\right)  \mu^{4-D} + \delta_{[D]4} \; \mu^{4-D}\Bigg]\int\frac{d^D q}{(2\pi)^D}\frac{1}{\left[q^{2}-\Delta^{2}\right]^2} 
\end{eqnarray}
where the divergent integral evaluates to
\begin{equation} \label{eqn:int_gamma2}
    \int\frac{d^D q}{(2\pi)^D}\frac{1}{\left[q^{2}-\Delta^{2}\right]^2} = \frac{i}{(4 \pi)^{D/2}} \frac{\Gamma(2-D/2)}{\Gamma(2)}(\Delta^2)^{D/2-2} \: .
\end{equation}
Making use of (\ref{eqn:int_gamma2}) in (\ref{eqn:i2_1}) before summing it up with (\ref{eqn:i1_2}) and then taking the ($D\to4$) limit for $I_{1}+I_{2}$ leads to 
\begin{equation} \label{eqn:i1_i2}
    (I_{1}+I_{2})_{D\to4}=\frac{i}{16\pi^2}\: \left[p^{\mu}p^{\nu}(2x-1)^2-p^2\,g^{\mu\nu}(2x-1)^2+p^2\,g^{\mu\nu}(2x-1)\right]\log \frac{\mu^2}{\Delta^2}
\end{equation}
where $\Delta^2$ is as given in (\ref{eqn:feyn_delta}). The next step is to plug (\ref{eqn:i1_i2}) and (\ref{eqn:i1_3}) into

\begin{equation}
    i\Pi^{\mu\nu}_{\ref{fig:sqed_photon}a+\ref{fig:sqed_photon}b}\,(p)=e^2\int\limits_{0}^{1} dx\;\Big\{(I_1)_{D\to2}+(I_{1}+I_{2})_{D\to4}\Big\}
\end{equation}
which can be put into the following final form via the change of variable $y=x-1/2$
 
\def\at{
  \left.
  \vphantom{\int}
  \right|
}

\begin{equation}
    i\Pi^{\mu\nu}_{\ref{fig:sqed_photon}a+\ref{fig:sqed_photon}b}\,(p)= -\:\frac{ i e^2 }{16 \pi^2}\, g^{\mu \nu}\; \Lambda_\wp^2 \: -\frac{ i e^2 }{4\pi^2}\: [p^2g^{\mu\nu}-p^{\mu} p^{\nu}]\; \int\limits_{-1/2}^{1/2} dy\Bigg(y^2\: \log \Bigg[\frac{\mu^2}{p^2(y^2-\frac{1}{4})+m^2}\Bigg]\Bigg) \; .
\end{equation}

\subsubsection*{Scalar propagator in Scalar Electrodynamics}

One-loop corrections to the scalar propagator in SQED is depicted in FIG. \ref{fig:sqed_scalar}. The amplitude of the self energy correction in FIG. \ref{fig:sqed_scalar} (a) reads
\begin{equation}
    i \Pi_{\ref{fig:sqed_scalar}a}(p)= -e^2\int\frac{d^4 k}{(2\pi)^4}\, \frac{P_{\mu\nu}(k) (k+2p)^{\mu} (k+2p)^{\nu}}{k^2 \, [(k+p)^2-m^2]}
\end{equation}
where 
\begin{equation}
    P_{\mu\nu}(k)= g_{\mu\nu}-\frac{k_{\mu}k_{\nu}}{k^2}
\end{equation}
is the projector of the photon propagator in Lorenz gauge which satisfies $k^{\mu}P_{\mu\nu}=0$
and $k^{\nu}P_{\mu\nu}=0$. Using these two relations we obtain
\begin{equation} 
    i \Pi_{\ref{fig:sqed_scalar}a}(p)= -4 e^2 p^2 \left(1-\frac{1}{D}\right)\int\frac{d^4 k}{(2\pi)^4}\, \frac{1}{k^2 \, [(k+p)^2-m^2]} \, .
\end{equation}
First, Feynman-parameterizing the denominator and then applying the regularization (\ref{int-Dd}), one obtains

\begin{equation} \label{eqn:scalar_self_sqed}
    i \Pi_{\ref{fig:sqed_scalar}a}(p)= -4 e^2 p^2 \int\limits_{0}^{1} dx\left(1-\frac{1}{D}\right)\, \Bigg[\left(\delta_{[D]0}+\delta_{[D]2}\right) \mu^{4-D} + \delta_{[D]4}\; \mu^{4-D}\Bigg]\int\frac{d^D q}{(2\pi)^D}\, \frac{1}{[q^2-\Delta^2]^2} \, ,
\end{equation}
where the momentum integral is again given by (\ref{eqn:int_gamma2}) and $\Delta^2=p^2(x^2-x)+m^2x$. Since the power of the divergent integral is $n=2$ we need only consider the ($D\to4$) limit for (\ref{eqn:scalar_self_sqed}) while applying the new regularization. This results in 

\begin{equation}
    i \Pi_{\ref{fig:sqed_scalar}a}(p)= \frac{-3\, i\, e^2 \, p^2}{16\pi^2}\, \Bigg\{\log \;  \Bigg(\frac{\mu^2}{m^2-p^2}\Bigg) - \frac{m^2}{p^2}\, \log \Bigg(\frac{m^2}{m^2-p^2}\Bigg) + \frac{4}{3}\Bigg\} \; ,
\end{equation}
where  $\overline{MS}$ subtraction scheme is applied again.

The amplitude for the scalar-scalar tadpole correction in FIG. \ref{fig:sqed_scalar} (b) is given by 

\begin{equation}
    i \Pi_{\ref{fig:sqed_scalar}b}= \lambda \int\frac{d^4 k}{(2\pi)^4}\, \frac{1}{k^2 -m^2}
\end{equation}
where $m$  is the mass of the charged scalar. Taking note of the fact that the power of the denominator of the propagator is $n=1$, the regularization (\ref{int-Dd}) yields

\begin{equation}
    i \Pi_{\ref{fig:sqed_scalar}b}= \lambda\; \Bigg[\frac{1}{8 \pi}\left(\delta_{[D]0}+\delta_{[D]2}\right)\Lambda_\wp^{2} \: \: \mu^{2-D} + \delta_{[D]4}\;\mu^{4-D}\Bigg] \int\frac{d^D k}{(2\pi)^D}\, \frac{1}{k^2 -m^2}
\end{equation}
where again the integral is given by 
\begin{equation}\label{eqn:int_gamma1_1}
    \int\frac{d^D k}{(2\pi)^D}\:\frac{1}{k^{2}-m^{2}} = \frac{-i}{(4 \pi)^{D/2}} \frac{\Gamma(1-D/2)}{\Gamma(1)}(m^2)^{D/2-1} \: .
\end{equation}

Since $n=1$, we need to take both ($D\to2$) and ($D\to4$) limits in performing the detached regularization on $\Pi_{\ref{fig:sqed_scalar}b}$. The analytical continuation to $D=2$ results in

\begin{equation}
    (i \Pi_{\ref{fig:sqed_scalar}b})_{D\to2}= \frac{- i\, \lambda}{32 \pi^2}\, \Lambda_\wp^2 \, \log \;  \Bigg(\frac{\mu^2}{m^2}\Bigg)  \;, 
\end{equation}
while analytical continuation to $D=4$ gives

\begin{equation}
    (i \Pi_{\ref{fig:sqed_scalar}b})_{D\to4}= \frac{ i\, \lambda}{16\pi^2}\, m^2 \, \Bigg\{\log \;  \Bigg(\frac{\mu^2}{m^2}\Bigg) +1\Bigg\} \;
\end{equation}
where  $\overline{MS}$ subtraction scheme is applied in both $D=2$ and $D=4$ cases.
Putting it all together, the new regularization (\ref{int-Dd}) and $\overline{MS}$ subtraction scheme give the  final result 

\begin{equation}
    i \Pi_{\ref{fig:sqed_scalar}b}= \frac{- i\, \lambda}{32 \pi^2}\, \Lambda_\wp^2 \, \log \;  \Bigg(\frac{\mu^2}{m^2}\Bigg)  \; +\frac{ i\, \lambda}{16\pi^2}\, m^2 \, \Bigg\{\log \;  \Bigg(\frac{\mu^2}{m^2}\Bigg) +1\Bigg\} \;.
\end{equation}

The last diagram that we will examine which contributes to the scalar propagator at one-loop level is the scalar-photon tadpole given in FIG. \ref{fig:sqed_scalar} (c). The amplitude for this is

\begin{equation} \label{eqn:sqed_sptad}
    i \Pi_{\ref{fig:sqed_scalar}c}= 2\; e^2\; \int\frac{d^4 k}{(2\pi)^4}\, \frac{g^{\mu \nu} P_{\mu \nu}(k)}{k^2 -m_{\gamma}^{2}}
\end{equation}
where $P_{\mu \nu}(k)$ is again given by (\ref{eqn:photon_prop}) and whose contraction with the metric is 
\begin{equation}
    g^{\mu \nu} P_{\mu \nu}(k)=D-1 \:.
\end{equation}
This makes (\ref{eqn:sqed_sptad}) 
\begin{equation} \label{eqn:sqed_sptad1}
    i \Pi_{\ref{fig:sqed_scalar}c}= 2\; e^2\; (D-1)\int\frac{d^4 k}{(2\pi)^4}\, \frac{1}{k^2 -m_{\gamma}^{2}} \:.
\end{equation}
The detached regularization applied to (\ref{eqn:sqed_sptad1}) yields

\begin{equation} 
    i \Pi_{\ref{fig:sqed_scalar}c}= 2\; e^2\; (D-1)\Bigg[\frac{1}{8 \pi}\left(\delta_{[D]0}+\delta_{[D]2}\right) \Lambda_\wp^{2} \: \: \mu^{2-D} + \delta_{[D]4} \; \mu^{4-D}\Bigg]\int\frac{d^D k}{(2\pi)^D}\, \frac{1}{k^2 -m_{\gamma}^{2}} 
\end{equation}
in which the momentum integral amounts to
\begin{equation}\label{eqn:int_gamma1_2}
    \int\frac{d^D k}{(2\pi)^D}\:\frac{1}{k^2 -m_{\gamma}^{2}} = \frac{-i}{(4 \pi)^{D/2}} \frac{\Gamma(1-D/2)}{\Gamma(1)}\; (m_{\gamma}^{2})^{D/2-1} \: .
\end{equation}
Since the power of the denominator is $n=1$, both ($D\to2$) and ($D\to4$) limits should be considered in performing the detached regularization.
The ($D\to2$) limit yields 

\begin{equation}
    (i \Pi_{\ref{fig:sqed_scalar}c})_{D\to2}= \frac{ i\, e^2}{8\pi^2}\, \Lambda_\wp^2 \; -\frac{ i\, e^2}{16\pi^2}\, \Lambda_\wp^2 \log \Bigg( \frac{\mu^2}{m_{\gamma}^{2}}\Bigg)  \; 
\end{equation}
whereas the ($D\to4$) limit yields no contribution since it basically amounts to an amplitude which is proportional to the square of the photon mass.
Therefore the  resultant amplitude in the  $\overline{MS}$ subtraction scheme becomes

\begin{equation}
    i \Pi_{\ref{fig:sqed_scalar}c}= \frac{ i\, e^2}{8\pi^2}\, \Lambda_\wp^2 \; - \frac{ i\, e^2}{16\pi^2}\, \Lambda_\wp^2 \log \Bigg( \frac{\mu^2}{m_{\gamma}^{2}}\Bigg) \;.
\end{equation}

\subsection*{Appendix B: Spinor Electrodynamics }
\subsubsection*{Vacuum polarization in Spinor Electrodynamics}
According to FIG. \ref{fig:qed} (a), the amplitude of vacuum polarization is given by
\begin{equation}
i\Pi^{\mu \nu}_{\ref{fig:qed}a}(p)=-e^2\int \frac{d^4 k}{(2\pi)^4}\frac{Tr\left[\gamma^{\mu}(\slashed{k}+\slashed{p}+m_{f})\gamma^{\nu} (\slashed{k}+m_{f})\right]}{\left[k^2-m_{f}^{2}\right]\left[(k+p)^2-m_{f}^{2}\right]} \; .
\end{equation}
The trace in the numerator amounts to
\begin{equation}
    Tr\left[\gamma^{\mu}(\slashed{k}+\slashed{p}+m_{f})\gamma^{\nu}(\slashed{k}+m_{f})\right]= D\left[(k^{\mu}+p^{\mu})k^{\nu}+(k^{\nu}+p^{\nu})k^{\mu}+g^{\mu\nu}(m_{f}^{2}-[k+p]\cdot k)\right]\:.
\end{equation}
Applying the Feynman parametrization one arrives at
\begin{eqnarray}\label{eqn:qed_vacuum}
    i\Pi^{\mu \nu}_{\ref{fig:qed}a}(p)=-e^2\: D \int\limits_{0}^{1} dx\;\int\frac{d^4 q}{(2\pi)^4}\,\left\{\left[\frac{2}{D} g^{\mu\nu}\frac{q^2}{\left[q^{2}-\Delta^{2}\right]^2} \; - \frac{g^{\mu\nu}}{q^{2}-\Delta^{2}}\right]+\frac{2(x-x^2)[p^2g^{\mu\nu}-p^{\mu} p^{\nu}]}{\left[q^{2}-\Delta^{2}\right]^2}\right\} 
\end{eqnarray}
where $\Delta^2=p^2(x^2-x)+m_{f}^2$. The first term inside the square brackets in (\ref{eqn:qed_vacuum}) can be decomposed into two irreducible integrals the result of which is 
\begin{equation}
    i\Pi^{\mu \nu}_{\ref{fig:qed}a}\,(p)=-e^2\int\limits_{0}^{1} dx\;\Big\{ D \: I_1+ D \: I_2\Big\}
\end{equation}
where
\begin{equation}
    I_{1}=g^{\mu\nu}\left(\frac{2}{D}-1\right) \int\frac{d^4 q}{(2\pi)^4}\,\frac{1}{q^{2}-\Delta^{2}}
\end{equation}
and 
\begin{equation}
    I_{2}=\frac{2}{D}\; g^{\mu\nu}\Delta^{2} \int\frac{d^4 q}{(2\pi)^4}\,\frac{1}{\left[q^{2}-\Delta^{2}\right]^2}+2(x-x^2)[p^2g^{\mu\nu}-p^{\mu} p^{\nu}]\int\frac{d^4 q}{(2\pi)^4}\,\frac{1}{\left[q^{2}-\Delta^{2}\right]^2}\:.
\end{equation}
First, we apply the regularization (\ref{int-Dd}) to $I_1$, paying attention to the fact that the power of the denominator is $n=1$. This means that we need to consider not only the analytical continuation to $D=4$ but also to $D=2$. The regularization applied to $I_1$ reads

\begin{equation}\label{eqn:i1_qed}
    I_1=g^{\mu\nu}\left(\frac{2}{D}-1\right)\;\Bigg[\frac{1}{8 \pi}\left(\delta_{[D]0}+\delta_{[D]2}\right) \Lambda_\wp^{2} \: \: \mu^{2-D} + \delta_{[D]4} \; \mu^{4-D}\Bigg] \;\int\frac{d^D q}{(2\pi)^D}\:\frac{1}{q^{2}-\Delta^{2}}
\end{equation}
where the momentum integral is again given by (\ref{eqn:int_gamma1}). In the $D\to2$ limit (\ref{eqn:i1_qed}) yields

\begin{equation}\label{eqn:i1_qed1}
    (I_1)_{D\to2}=-ig^{\mu\nu}\frac{\Lambda_\wp^{2}}{32\pi^2}
\end{equation}
where we  used the $\overline{MS}$ subtraction scheme. 

Now, let's evaluate $I_2$. The power of the denominator of the divergent integrals in $I_2$ is $n=2$, therefore the new regularization scheme (\ref{int-Dd}) applied to it reads

\begin{eqnarray}\label{eqn:i2_qed}
    I_2&=&g^{\mu\nu}\frac{2}{D}\, \Delta^2 \Bigg[\left(\delta_{[D]0}+\delta_{[D]2}\right) \mu^{4-D} + \delta_{[D]4} \; \mu^{4-D}\Bigg]\:\int\frac{d^D q}{(2\pi)^D}\,\frac{1}{\left[q^{2}-\Delta^{2}\right]^2} \\ \nonumber
        &+&2(x-x^2)[p^2g^{\mu\nu}-p^{\mu} p^{\nu}]\Bigg[\left(\delta_{[D]0}+\delta_{[D]2}\right) \mu^{4-D} + \delta_{[D]4} \; \mu^{4-D}\Bigg]\int\frac{d^D q}{(2\pi)^D}\frac{1}{\left[q^{2}-\Delta^{2}\right]^2} 
\end{eqnarray}
where the momentum integral is given by (\ref{eqn:int_gamma2}). Next we sum (\ref{eqn:i2_qed}) with (\ref{eqn:i1_qed}) and then take the ($D\to4$) limit for $I_{1}+I_{2}$. This leads to 
\begin{equation} \label{eqn:i1_i2_qed}
    (I_{1}+I_{2})_{D\to4}=\frac{i}{8\pi^2}\: \left[p^{\mu}p^{\nu}-p^2\,g^{\mu\nu}\right](x^2-x)\log \frac{\mu^2}{\Delta^2}
\end{equation}
where $\Delta^2=p^2(x^2-x)+m_{f}^2$. The next step is to plug (\ref{eqn:i1_i2_qed}) and (\ref{eqn:i1_qed1}) into
\begin{equation}
    i\Pi^{\mu \nu}_{\ref{fig:qed}a}\,(p)=-e^2\int\limits_{0}^{1} dx\;\Big\{ [D \: I_1]_{D\to2}+ [D \: (I_1+I_2)]_{D\to4}\Big\}
\end{equation}
which results in the  regularized amplitude

\begin{equation}
i\Pi^{\mu \nu}_{\ref{fig:qed}a}(p)=\frac{ie^2}{16\pi^2}g^{\mu\nu} \Lambda_\wp^{2} -\frac{ i e^2 }{2\pi^2}\: [p^2g^{\mu\nu}-p^{\mu} p^{\nu}]\; \int\limits_{0}^{1} dx \left\{\: (x-x^2)\:\left(- \:\,\frac{1}{2} + \log \left[\frac{\mu^2}{p^2(x^2-x)+m_{f}^{2}}\right]\right)\right\} \; ,
\end{equation}
where the second term is the finite transverse term.

\subsubsection*{Fermion propagator in Spinor Electrodynamics}
The correction to the fermion propagator is given by the diagram in FIG. \ref{fig:qed} (b) which corresponds to
\begin{equation}
-i\Sigma_{\ref{fig:qed}b}(p)=-e^2\int \frac{d^4 k}{(2\pi)^4}\frac{\gamma^{\mu}(\slashed{k}+m_{f})\gamma_{\mu}}{\Big[k^2-m_{f}^2\Big]\Big[(p-k)^2-m_{\gamma}^2\Big]} \; 
\end{equation}
wherein the numerator equals 
\begin{equation}
    \gamma^{\mu}(\slashed{k}+m_{f})\gamma_{\mu}=(2-D)\slashed{k}+m_{f}D\; .
\end{equation}
The next step is to Feynman parameterize the denominator and shift the loop momenta accordingly. The outcome of this step is
\begin{equation}
    -i\Sigma_{\ref{fig:qed}b}(p)=e^2\int\limits_{0}^{1} dx \left\{\left[(D-2)\slashed{p}x-m_{f}D\right]\int \frac{d^4 q}{(2\pi)^4}\frac{1}{\Big[q^2-\Delta^2\Big]^2} \right\}\; 
\end{equation}
where the terms linear in the shifted loop momenta are dropped since they evaluate to zero. Next, the regularization (\ref{int-Dd}) is applied 

\begin{equation}
    -i\Sigma_{\ref{fig:qed}b}(p)=e^2\int\limits_{0}^{1} dx \left\{\left[(D-2)\slashed{p}x-m_{f}D\right]\Bigg[\left(\delta_{[D]0}+\delta_{[D]2}\right) \mu^{4-D} + \delta_{[D]4} \; \mu^{4-D}\Bigg]\int\frac{d^D q}{(2\pi)^D}\, \frac{1}{[q^2-\Delta^2]^2}  \right\}\; 
\end{equation}
in which the momentum integral is again given by (\ref{eqn:int_gamma2})  and $\Delta^2=p^2(x^2-x)+xm_{\gamma}^2+(1-x)m_{f}^2$. Since the power of the divergent integral is $n=2$ we need only consider the ($D\to4$) limit. This results in 
\begin{equation}
-i\Sigma_{\ref{fig:qed}b}(p)= \frac{ i e^2 }{16\pi^2}\:\int\limits_{0}^{1} dx \left[2\, m_{f} - 2 x\, \slashed{p}\right]\; - \;  \frac{ i e^2 }{16\pi^2}\:\int\limits_{0}^{1} dx \left\{\left[4\, m_{f} - 2 x\, \slashed{p}\right]\;\log \left[\frac{\mu^2}{p^2(x^2-x) + x \, m_{\gamma}^2 + (1-x)\, m_{f}^{2}}\right]\right\} 
\end{equation}
via the $\overline{MS}$ subtraction scheme.

\subsubsection*{One-loop vertex correction in Spinor Electrodynamics}
The correction to the fermion-photon vertex in spinor QED is given by the diagram FIG. \ref{fig:qed} (c) which is  
\begin{equation}
-i e \Gamma^{\mu}_{\ref{fig:qed}c}(p)=-e^3\int \frac{d^4 k}{(2\pi)^4}\frac{\gamma_{\nu}(\slashed{p}^{'}-\slashed{k}+m_{f})\gamma^{\mu} (\slashed{p}-\slashed{k}+m_{f})\gamma^{\nu}}{k^2 \, \Big[(p-k)^2-m_{f}^2\Big]\Big[(p^{'}-k)^2-m_{f}^2\Big]} \;.
\end{equation}
The first step is Feynman parameterizing the denominator. This leads to
\begin{equation}
    -i e \Gamma^{\mu}_{\ref{fig:qed}c}(p)=-2\,e^3 \int\limits_{0}^{1} dx \, \int\limits_{0}^{1-x} dy \int \frac{d^4 q}{(2\pi)^4}\frac{\gamma_{\nu}[\slashed{p}^{'}(1-y)-\slashed{p}x-\slashed{q}+m_{f}]\gamma^{\mu} [\slashed{p}(1-x)-\slashed{p}^{'}y-\slashed{q}+m_{f})\gamma^{\nu}}{\left[q^2-\Delta^2\right]^3} \; \label{eqn:qed_vertex}
\end{equation}
where
\begin{equation} 
    \Delta^{2}=\, m_{f}^2 \, (x+y) + p^2 \, (x^{2}-x) + {p^{'}}^{2} \, (y^{2}-y)\,+2\,p \cdot p^{'} (xy) \: \label{eqn:delta_vertex}
\end{equation}
and $q$ is the shifted loop momentum. The numerator of (\ref{eqn:qed_vertex}) can be written as
\begin{equation}
    \gamma_{\nu}[\slashed{p}^{'}(1-y)-\slashed{p}x-\slashed{q}+m_{f}]\gamma^{\mu} [\slashed{p}(1-x)-\slashed{p}^{'}y-\slashed{q}+m_{f})\gamma^{\nu} =\gamma_{\nu}\slashed{q}\gamma^{\mu}\slashed{q}\gamma^{\nu}+\tilde{N}^{\mu}+ {\rm \ (terms\ linear\ in \ }q)
\end{equation}
where
\begin{equation}
    \tilde{N}^{\mu}= \gamma_{\nu}\left[\slashed{p}^{'}(1-y)-\slashed{p}x+m_{f}\right]\gamma^{\mu} \left[\slashed{p}(1-x)-\slashed{p}^{'}y+m_{f}\right]\gamma^{\nu} \: .
    \label{eqn:n_mu_tilde}
\end{equation}
The gamma matrix algebra in $D$ dimensions applied to (\ref{eqn:qed_vertex}) leads to 
\begin{equation}
    -i e \Gamma^{\mu}_{\ref{fig:qed}c}(p)=-2\,e^3 \int\limits_{0}^{1} dx \, \int\limits_{0}^{1-x} dy \left\{\frac{(2-D)^2}{D}\gamma^{\mu}\int \frac{d^4 q}{(2\pi)^4}\frac{q^2}{\left[q^2-\Delta^2\right]^3}+\tilde{N}^{\mu}\int \frac{d^4 q}{(2\pi)^4}\frac{1}{\left[q^2-\Delta^2\right]^3}\right\}
     \; \label{eqn:qed_vertex1}
\end{equation}
wherein the first integral can be reduced to two irreducible ones which are
\begin{equation}
    I_1=\frac{(2-D)^2}{D}\gamma^{\mu}\int \frac{d^4 q}{(2\pi)^4}\frac{1}{\left[q^2-\Delta^2\right]^2}
\end{equation}
and
\begin{equation}
    I_2=\frac{(2-D)^2}{D}\gamma^{\mu}\Delta^2 \int \frac{d^4 q}{(2\pi)^4}\frac{1}{\left[q^2-\Delta^2\right]^3}
\end{equation}
and the last integral in (\ref{eqn:qed_vertex1})
\begin{equation}
    I_3=\tilde{N}^{\mu}\int \frac{d^4 q}{(2\pi)^4}\frac{1}{\left[q^2-\Delta^2\right]^3}
\end{equation}
is kept intact.
The first integral is regularized via (\ref{int-Dd}) taking note of the fact that $n=2$ therefore we only  make the analytical continuation $D\to 4$ which leads to
\begin{equation}
    I_1=\frac{ i \gamma^{\mu} }{16\pi^2}\left(-2\,+\,\log \frac{\mu^2}{\Delta^{2}}\right)\:.
\end{equation}
The other two integrals $I_2$ and $I_3$ are already convergent. Therefore there is no need to regularize them. They amount to 
\begin{equation}
    I_2= \frac{ -i \gamma^{\mu} }{32\pi^2}
\end{equation}
and 
\begin{equation}
    I_3= \frac{ -i \tilde{N}^{\mu} }{32\pi^{2}\Delta^2}\:.
\end{equation}
Putting it all together as 
\begin{equation}
    -i e \Gamma^{\mu}_{\ref{fig:qed}c}(p)=-2\,e^3 \int\limits_{0}^{1} dx \, \int\limits_{0}^{1-x} dy \left\{I_1 + I_2 + I_3 \right\}
     \;
\end{equation}
the regularized form of the vertex correction becomes
\begin{equation}
-i e \Gamma^{\mu}_{\ref{fig:qed}c}(p)=-2\,e^3 \int\limits_{0}^{1} dx \, \int\limits_{0}^{1-x} dy  \left\{\frac{ i \gamma^{\mu} }{16\pi^2}\left(-\frac{5}{2}\,+\,\log \frac{\mu^2}{\Delta^{2}}\right)\,-\, \frac{ i \tilde{N}^{\mu} }{32\,\pi^2\, \Delta^{2}}\right\}\:,
\end{equation} 
where $\Delta^2$ is given by (\ref{eqn:delta_vertex}) and $\tilde{N}^{\mu}$ by (\ref{eqn:n_mu_tilde}).

\end{document}